\documentclass[lettersize,journal]{IEEEtran}
\usepackage{amsmath,amsfonts}
\usepackage{algorithmic}
\usepackage{algorithm}
\usepackage{array}
\usepackage[caption=false,font=normalsize,labelfont=sf,textfont=sf]{subfig}
\usepackage{xcolor}
\usepackage[hidelinks]{hyperref}
\usepackage{textcomp}
\usepackage{stfloats}
\usepackage[
  backend=biber,
  style=numeric-comp,
  sorting=none,
  sortcites=true,
  doi=false,
  url=false,
  isbn=false,
  eprint=false,
]{biblatex}
\AtEveryBibitem{
  \clearfield{publisher}
  \clearlist{publisher}
  \clearfield{location}
  \clearlist{location}
  \clearfield{note}
  \clearfield{language}
}

\usepackage{url}
\usepackage{verbatim}
\usepackage{graphicx}
\usepackage{caption}
\newcommand{\bI}{\boldsymbol{I}}
\newcommand{\bQ}{\boldsymbol{Q}}
\newcommand{\bIp}{\boldsymbol{I_p}}
\newcommand{\bQp}{\boldsymbol{Q_p}}
\newcommand{\ibw}{\mathrm{IBW}}
\newcommand{\Ex}[1]{\langle{ #1 }\rangle}
\newcommand{\snr}{\mathrm{SNR}}

\begin{document}

\title{Quantum-limited Optical Vector Analysis}

\author{Karthik Dasigi, Pavel A. Dmitriev, Kah Jen Wo, Fumiya Hanamura, Lingda Kong and Steven Touzard
\thanks{All authors are with the Center for Quantum Technologies, Singapore. P.D, F.H and S.T are also a part of the Department of Material Science and Engineering, NUS, Singapore. S.T is also affiliated with the Department of Physics, NUS, Singapore and Majulab, CNRS-UNS-NUS-NTU International Joint Research Unit UMI 3654, Singapore.}
}

\maketitle

\begin{abstract}
Optical Vector Analysers (OVA) are critical for emerging technologies such as integrated photonics and optical positioning. 
The sensitivity of existing OVAs falls orders of magnitude short of the Standard Quantum Limit (SQL).
Approaching the SQL while acquiring a wide spectrum allows accurate measurements of targets that are fragile, non-linear, or that scatter most of the probe light away. In this paper we propose and implement an OVA scheme capable of measuring at the SQL, based on frequency-swept interferometry and balanced heterodyne detection. 
We develop a quantum-mechanical formalism that quantifies the trade-offs between frequency resolution, sweep speed, and sensitivity in the presence of interferometric phase-noise.
This analysis enables our experimental demonstration of near-SQL performance with standard scanning lasers and straightforward phase-noise mitigation techniques.
We leverage our high sensitivity to characterize ring resonators with signal-to-noise ratio $>$1 at optical powers as low as 20\,fW.
\end{abstract}


\section{Introduction}
\label{sec:1}
Characterizing a physical system requires stimulating it with a known signal and observing its response. If the system is linear, a stimulus at a given frequency generates a response at the same frequency with a modified amplitude and phase. Scanning the frequency of the stimulus then provides the full response function of the system. For optical systems, this method is known as Optical Vector Analysis (OVA) and is paramount to the development of photonic integrated circuits \cite{canavesi_polarization-_2009, mckinnon_extracting_2009, pan_ultrahigh-resolution_2017}, navigation with LIDAR \cite{kuse_frequency-modulated_2019, shi_frequency-comb-linearized_2024}, and high resolution microscopy \cite{referee_cite} with Optical Coherent Tomography (OCT) \cite{huang_optical_1991, chen_high_2022}. 

Despite their broad applicability, much of the potential of OVAs remains untapped because the sensitivity of leading implementations \cite{cui_distinguishing_2024, xue_accuracy_2014, pan_ultrahigh-resolution_2017, bao_digitally_2015, gotti_comb-locked_2020, shi_frequency-comb-linearized_2024, qing_optical_2019,Wang2025SelfCalibratedCO} lies orders of magnitude away from the Standard Quantum Limit (SQL) \cite{yuenNoiseHomodyneHeterodyne1983}. 
In contrast, fixed-frequency measurements already attain sensitivities at the SQL or beyond, for example to sense gravitational waves \cite{ganapathyBroadbandQuantumEnhancement2023}.
Combining SQL sensitivity with swept-frequency operation is crucial for the characterization of devices that operate only at low power, such as those exploiting extreme nonlinearities in $\mathrm{BaTiO_3}$ \cite{cao_barium_2021}, doped systems \cite{xieScalableMicrowavetoopticalTransducers2025}, or other emerging material platforms \cite{JelenaEO}.
Integrated photonic devices also benefit from low power characterization, as the tight field confinement results in undesirable effects, such as the thermo-optic effect \cite{grayThermoOpticMultistabilityRelaxation2020} and the photorefractive effect \cite{mondainPhotorefractiveEffectLiNbO32020}.
By enabling efficient low power characterization of these devices, quantum-limited OVAs would substantially accelerate the fields of optical computing \cite{10909496, gao_all-optical_2025}, quantum computing \cite{psiquantumteamManufacturablePlatformPhotonic2025, budingerAllopticalQuantumComputing2024,AghaeeRad2025}, and quantum networking \cite{xieScalableMicrowavetoopticalTransducers2025}.






In the highest-sensitivity OVA reported to date \cite{Wang2025SelfCalibratedCO}, the probe signal from the Device Under Test (DUT) is mixed with a separate reference Local Oscillator (LO) prior to detection, resulting in an amplified photocurrent \cite{yuenNoiseHomodyneHeterodyne1983}. While this amplification provides increased sensitivity, this measurement scheme, known as coherent OVA, is susceptible to phase-noise \cite{qing_optical_2019,Wang2025SelfCalibratedCO}. This drawback is mitigated in incoherent OVAs, where the LO co-propagates with the probe \cite{cui_distinguishing_2024, xue_accuracy_2014, pan_ultrahigh-resolution_2017, bao_digitally_2015, gotti_comb-locked_2020, shi_frequency-comb-linearized_2024}. However, in this latter scheme, any broadband attenuation leads to reduced sensitivity due to an insufficiently large LO amplitude. Despite progress in both approaches, the lowest reported resolved power of 10\,nW \cite{Wang2025SelfCalibratedCO} remains several orders of magnitude far from the quantum-limited regime we predict. This gap points to an incomplete understanding of the fundamental sensitivity limits of OVAs. 

Here, we provide the missing analysis and experimentally demonstrate that it enables sensitivities approaching the quantum limit using a simple coherent OVA scheme. We present a complete quantum-theoretical treatment of the dependence of the minimum resolvable power on sweep speed and frequency resolution. This analysis indicates that phase-noise restricts the choices of sweeping parameters for which the sensitivity remains quantum-limited. We experimentally investigate the range of suitable sweeping parameters required to achieve SQL sensitivity in a simple scanning laser interferometer equipped with balanced detection. 
By analyzing and mitigating the phase-noise of our setup, we expand this range and characterize narrow LNOI ring resonators (quality factor $\mathrm{Q}>10^6$), with powers as low as 20\,fW. 
This minimum resolvable power corresponds to an improvement of six orders of magnitude over the current state of the art \cite{Wang2025SelfCalibratedCO}.

\section{Measurement Setup and Theory}
\label{sec:2}
\begin{figure*}[!t]
\makebox[\textwidth][c]{
\includegraphics{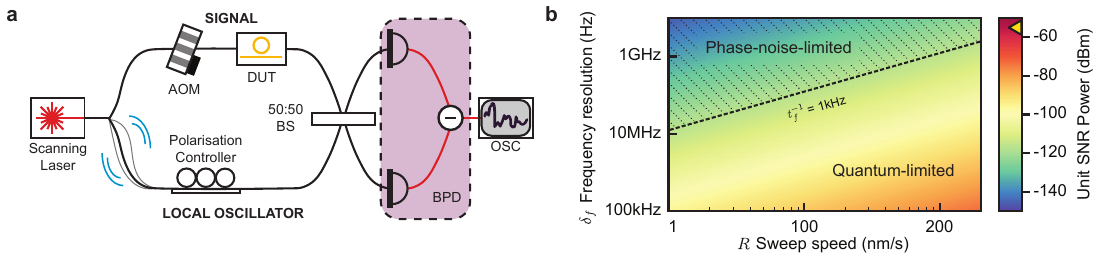}}
\caption{Schematic of the measurement setup and fundamental limits. (a) The Device Under Test (DUT) is characterized with a balanced heterodyne measurement, performed with a free-running Mach-Zehnder Interferometer (MZI) comprising a scanning laser, a Polarisation Controller, an Acousto-Optic Modulator (AOM), a beamsplitter (BS) and a Balanced Photo-Detector (BPD). The free-running MZI is sensitive to vibrations and thermal expansion, represented by the grey lines in the LO path. (b) The minimum measurable power at $\lambda_0$\,=\,1550\,nm as a function of the OVA's frequency resolution $\delta_f$ and sweep speed $R$. The dashed contour corresponds to a sweep time across $\delta_f$ that is one-tenth of the phase-noise timescale $t_f$\,=\,1ms. Below this contour, the measurement sensitivity is quantum-limited, whereas above, the sensitivity is phase-noise-limited. The yellow marker in the colorbar represents the predicted lowest measurable power from \cite{Wang2025SelfCalibratedCO}. The quantum efficiency $\eta$ used for this plot is that of our setup (Sec.~\ref{sec:4})}
\label{fig1}
\end{figure*}
Figure~\ref{fig1}a shows the setup of our quantum-limited OVA, which comprises only fiber components and mounted on a vibrationally isolated optical table. A scanning laser continuously emits light with an instantaneous frequency $\omega_0\approx 2\pi\times$200\,THz, while being chirped at a rate $R$ (nm/s). Most of the optical power is directed towards the LO branch with a 99:1 splitter. A fraction of the power is directed towards the probe branch, where its frequency is increased by $\omega_m = 2\pi\times$40\,MHz . After going through the DUT, the probe enters one port of a 50:50 beamsplitter (BS), while the LO enters the other port. Before entering the beamsplitter, the polarizations of the LO and probe are matched to maximize the sensitivity. The light at the output ports is detected on a Balanced Photodetector (BPD). Detailed information regarding the setup is presented in the supplement \cite{supp}.

The BPD output is a heterodyne beat signal at frequency $\omega_m$ \cite{yuenNoiseHomodyneHeterodyne1983, supp}. Measurement shots are obtained by demodulating this beat signal at $\omega_m$ over a time window $\tau$, corresponding to an Integration Bandwidth $\ibw=1/\tau$. For a coherent LO with large amplitude, the demodulated quadratures $\bI$ and $\bQ$ are Gaussian stochastic variables with statistics
\begin{gather}
    \Ex{\bI} = \eta\mathcal{R}\beta\Ex{\bIp},\quad
    \Ex{\bQ} =\eta\mathcal{R}\beta\Ex{\bQp},\label{eq:ExnIQ}\\
    \Ex{\Delta^2\bI} = \Ex{\Delta^2\bQ} = \frac{(\mathcal{R}\beta)^2\eta}{2\tau},\label{eq:varnIandnQ}
\end{gather}
where $\bIp$ and $\bQp$ are the quadratures of the probe signal at frequency $(\omega_0 + \omega_m)$ defined over a bandwidth $\ibw$, $\eta$ is the efficiency of the measurement scheme, $\mathcal{R}$ is the responsivity of the photodiodes and $\beta$ is the LO amplitude. The probe quadratures are defined with respect to the phase of the LO. 
Critically, as the average of the signal and its noise both scale with $\beta$, a large LO power renders the technical noise negligible, yielding a Signal-to-Noise Ratio ($\snr$)

\begin{equation}
    \snr=\frac{\Ex{\bI}^2+\Ex{\bQ}^2}{ \Ex{\Delta^2\bI} +\Ex{\Delta^2\bQ}}= \eta\,\frac{P}{\hbar\omega_0}\frac{1}{\mathrm{IBW}},\label{eqn:SNR_vs_IBW}
\end{equation}
where $\hbar$ is reduced Planck's constant and $P$ is the probe power.

As $\omega_0$ is swept at a rate $R$ (in nm/s), the time evolution of the measurement operators $\bI$ and $\bQ$ encodes the complex response of the DUT as a function of frequency. For a given sweep speed, the integration time $\tau$ sets the frequency resolution $\delta_f$ to
\begin{equation}
    \delta_f=\frac{R\,c}{\lambda_0^2}\,\tau=\frac{R\,c}{\lambda_0^2}\,\frac{1}{\ibw}\label{eqn:fres_vs_ibw},
\end{equation}
where $c$ is the speed of light and $\lambda_0 = c/\omega_0$ is the instantaneous wavelength. Since only the relative probe–LO phase is measured, any interferometric phase-noise corrupts the result. To avoid measurement degradation from phase-noise with characteristic timescale $t_f$, the sweep time over $\delta_f$ must satisfy
\begin{equation}
    \delta_f\,\frac{\lambda_0^2}{R\,c} \ll \,t_f,\label{eqn:fres_vs_phasenoise}
\end{equation}

Equations~\ref{eqn:SNR_vs_IBW}, \ref{eqn:fres_vs_ibw}, and \ref{eqn:fres_vs_phasenoise} define the fundamental limits of coherent OVA measurement employing balanced heterodyne detection. These limits are illustrated in Fig.~\ref{fig1}b for a wavelength of 1550\,nm. For a given sweep speed, one can sacrifice some frequency resolution in favor of a lower minimum measurable power. However, the phase-noise sets an upper bound on this trade-off. For modest sweep speeds, frequency resolutions and phase-noise timescales, the lowest measurable powers are well below the nW level. In general, the lowest detectable powers are obtained by sweeping as slowly as possible, up to the point where phase-noise can still be effectively mitigated.


\section{Mitigation of Classical Noise}
\label{sec:3}
\begin{figure}[!t]
\centering
\includegraphics{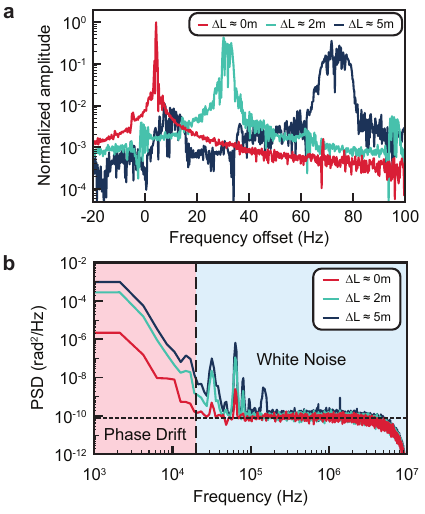}
\caption{Excess classical phase-noise. (a) Fourier transform of 3 signal traces, acquired at 200\,nm/s, with interferometer length mismatch of about 5, 2 and 0 meters. The delay introduces a frequency offset, and the peaks are broadened. (b) Power Spectral Density (PSD) of the phase-noise calculated with the Welch method. The 1/f noise dominates at low frequency (red background, left) and the quantum noise dominates after a few kHz (blue background, right). The dashed line corresponds to the predicted PSD \cite{supp}, computed for this probe power using the quantum efficiency of our setup $\eta$ (Sec.~\ref{sec:4}).}
\label{fig2}
\end{figure}

The two primary sources of phase noise in the setup are chirp rate fluctuations of the tunable laser and interferometric noise arising from differential path-length fluctuations. Of these, chirp rate noise is the dominant contribution, leading to an effective broadening of the laser linewidth during frequency sweeps \cite{shi_frequency-comb-linearized_2024, luo_wideband_2024}. Furthermore, fluctuations in the chirp rate couple to phase fluctuations in the free-running Mach–Zehnder interferometer (MZI), which generate additional phase-noise \cite{supp}. We mitigate this contribution by matching the optical path lengths of the interferometer arms \cite{QingLengthMatching}.

Fig.~\ref{fig2}a and Fig.~\ref{fig2}b show the Fourier transforms of the raw signal and the Power Spectral Densities (PSDs) of its phase during a wide frequency sweep, as the length mismatch of the interferometer arms is varied. As the length mismatch increases, the frequency peaks are offset and significantly broadened. The phase PSD reveals excess classical noise in the interferometer, which dominates over the white-noise background up to a knee frequency of approximately 20\,kHz. This noise is suppressed by more than an order of magnitude when the mismatch of the interferometer arm-length is under 1\,m.

To correct for the remaining classical noise, we divide the frequency range into windows of size $\Delta_f$ and detrend the phase within each window. The size $\Delta_f$ is chosen narrow enough for the phase to vary approximately linearly. This slow variation can be due to fluctuations either in sweep speed or in effective path length. The size of this window must be chosen carefully. If it is too narrow, the post-processing erases any phase feature with a large frequency range. If it is too wide, the phase-noise of the interferometer is not well corrected with a simple linear detrending. This correction matters most for slow sweep speeds, which are necessary to measure small powers.

\section{Sensitivity Characerization}
\label{sec:4}

Our analysis indicates that the measured phase-noise allows for a wide range of sweeping parameters for which the sensitivity is quantum-limited. At this sensitivity, the minimum measurable power is well below 1\,nW for the sweep speeds of our laser and for relevant frequency resolutions. To reach this quantum limit, we first verify that the LO power is sufficiently high as introduced in Sec.~\ref{sec:2}. We observe a quantum-noise-dominated noise floor when the LO is on while the probe arm is disconnected \cite{yuenNoiseHomodyneHeterodyne1983, supp}.

\begin{figure}[!t]
\centering
\includegraphics{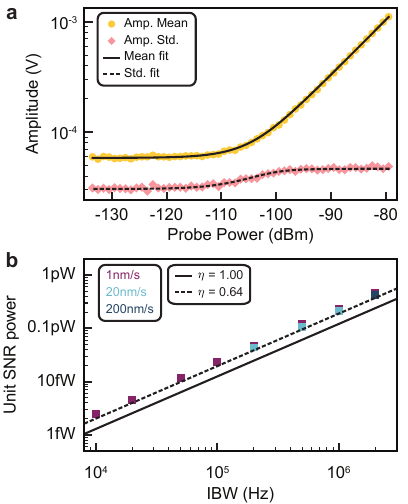}
\caption{Quantum efficiency and minimum power detectable. (a) Amplitude mean (squares) and standard deviation (diamonds) of the signal during a frequency sweep as a function of probe power ($\ibw$\,=\,200\,kHz). The mean and standard deviations are simultaneously fitted to a Rice distribution (solid and dashed lines) to extract the efficiency $\eta$\,=\,0.64. (b) Minimum average power detected with an SNR of at least 1 as a function of IBW. The lines correspond to the theoretical limit of zero phase-noise when $\eta$\,=\,1 and 0.64.}
\label{fig3}
\end{figure}

We assess the sensitivity, independently of the phase fluctuations, of our quantum-limited setup by measuring the amplitude response at the single-photon level during a wide frequency sweep. We acquire the amplitude of the beat signal and plot its mean and standard deviation as a function of the calibrated probe power in Fig.~\ref{fig3}a. At low powers, the resulting SNR differs strongly from simple photon detection due to the zero-point fluctuations of a coherent state. We simultaneously fit the mean and standard deviation of the amplitude with a Rician distribution, from which we deduce the efficiency of our measurement $\eta$\,=\,0.64. To report an efficiency near the quantum limit, the probe power down to -90\,dBm was calibrated using an optical power meter (Keysight N7748C), and lower powers were obtained via a calibrated variable attenuator (Santec OVA-100).

Finally, to quantify the overall sensitivity of the complete setup, we report the minimum probe power for which the complex amplitude of the signal can be reliably measured.
In Fig.~\ref{fig3}b, we plot the average probe power for which $\snr$\,=\,1 for different values of $R$. The unit-$\snr$ powers are slightly higher than those predicted by our measured $\eta$, due to the limitations of our detrending. 
Nevertheless, our measurement resolves the complex signal with approximately 1.6 photons of noise within a single integrated time bin.
This constant photon number implies a minimum power that scales linearly with bandwidth, yielding 2.4\,fW for a 10\,kHz bandwidth (at 1550\,nm).


\section{Application to Integrated Photonics}
\label{sec:5}
\begin{figure*}[!t]
\centering
\includegraphics{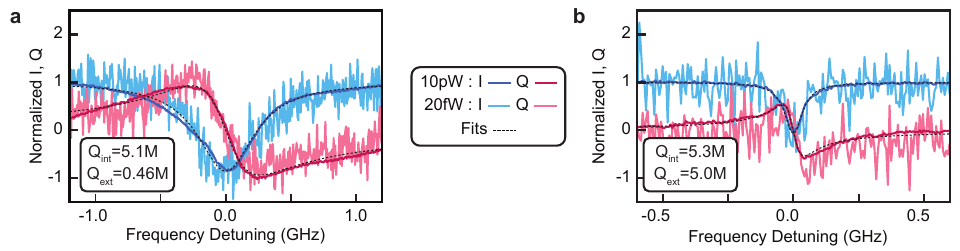}
\caption{Low-power characterisation of TFLN ring resonators without prior information. The IQ-space trajectory is normalised to the maximum transmission of the fit curve. I and Q display a constant amplitude and a large phase-roll for the over-coupled ring (a), while I and Q simultaneously go through 0 for the critically ring (b). The data is taken for moderate power (10\,pW) and low power (20\,fW), at a sweep speed of 1\,nm/s and an IBW of 20\,kHz. The I and Q quadratures are simultaneously fitted to the complex response of a harmonic oscillator, from which we extract the internal and external quality factors ($\mathrm{Q_{int}}$ and $\mathrm{Q_{ext}}$ respectively).}
\label{fig4}
\end{figure*}
Finally, we showcase our length-matched OVA in a key application, which consists in characterising photonic components without prior information. Here, our DUTs are ring resonators fabricated in thin-film Lithium Niobate (TFLN). We probe our ring resonators such that the output powers after the chip are far from our minimum detectable power ($P\approx$\,10\,pW) and close to it ($P\approx$\,20\,fW). These powers were measured using the same calibration process presented in the previous section. The on-chip power is 10\,dB higher than the power measured before the final beamplitter due to fibre-chip coupling loss. This on-chip power corresponds to an average intra-resonator photon number of 1 at the higher power and 0.01 at the lower power, ensuring that nonlinear effects remain negligible. With each frequency point integrated over an IBW of 20\,kHz, the total photon flux is also negligible, allowing slow nonlinear effects such as the photorefractive and thermo-optic responses to be safely ignored.

Fig.~\ref{fig4}a and Fig.~\ref{fig4}b show the demodulated IQ quadratures from single-shot measurements of an over-coupled and a critically coupled resonator respectively, demonstrating an $\snr>$\,1 at probe powers of 20\,fW. The quadratures are simultaneously fitted as a function of frequency, and display the characteristic simultaneous passage through 0 when critically coupled, and the characteristic constant amplitude and large phase-roll when over-coupled. These measurements demonstrate that our OVA operates close to our predicted quantum limit, and that it can measure powers that were previously orders of magnitude out of reach.

\section{Conclusion and Discussion}
\label{sec:6}
In conclusion, we have developed a quantum treatment to determine whether the sensitivity of OVAs can reach the fundamental quantum limit.
This analysis shows that a standard coherent OVA can operate near the quantum limit using only straightforward noise-mitigation techniques. 
In practice, reaching this limit enabled the accurate characterization of narrow LNOI ring resonators at optical powers down to 20\,fW, representing a six-orders-of-magnitude improvement over the current state of the art.

Beyond its improved sensitivity, our simple setup is directly compatible with previous works on improved frequency precision. In particular, the sweep speed of the scanning laser can be linearized using a calibrated interferometer \cite{puckett_422_2021,luo_wideband_2024}, a frequency comb \cite{gotti_comb-locked_2020,shi_frequency-comb-linearized_2024}, or by implementing single-sideband scanning via microwave modulation \cite{Wang2025SelfCalibratedCO}. The scanning range can also be scaled up by sequentially triggering scanning lasers over multiple bands \cite{luo_wideband_2024}. Finally, the setup is polarization-sensitive, as the heterodyne beat signal is present only when the local oscillator and probe share the same polarization.

\printbibliography

\end{document}